\documentclass[bibtex,twocolumn,prl,showpacs,floatfix,preprintnumbers,amsmath,amssymb,superscriptaddress,letter]{revtex4}

\usepackage{graphicx}
\usepackage{dcolumn}
\usepackage{color}

\newcommand{\numutonutau}{$\nu_{\mu}\rightarrow\nu_{\tau}$ }

\newcommand{\cs}{$\chi^{2}$}

\newcommand{\evsq}{${\rm eV}^{2}$}

\newcommand{\nutau}{$\nu_{\tau}$}
\newcommand{\dmsqtwo}{$|\Delta m^{2}_{32}|$}
\newcommand{\dmsq}{$|\Delta m^{2}|$}
\newcommand{\dmsqone}{$|\Delta m^{2}_{31}|$}

\newcommand{\sintwo}{$\rm sin^{2}(2\theta)$}

\begin{document}

\preprint{FERMILAB-PUB-08-168-E}
\preprint{BNL-xxxx}
\preprint{arXiv:hep-ex/0806.2237}
\title{Measurement of Neutrino Oscillations \\with the MINOS 
Detectors in the NuMI Beam}
%

\newcommand{\Cambridge}{Cavendish Laboratory, University of Cambridge, Madingley Road, Cambridge CB3 0HE, United Kingdom}
\newcommand{\FNAL}{Fermi National Accelerator Laboratory, Batavia, Illinois 60510, USA}
\newcommand{\RAL}{Rutherford Appleton Laboratory, Chilton, Didcot, Oxfordshire, OX11 0QX, United Kingdom}
\newcommand{\UCL}{Department of Physics and Astronomy, University College London, Gower Street, London WC1E 6BT, United Kingdom}
\newcommand{\Caltech}{Lauritsen Laboratory, California Institute of Technology, Pasadena, California 91125, USA}
\newcommand{\ANL}{Argonne National Laboratory, Argonne, Illinois 60439, USA}
\newcommand{\Athens}{Department of Physics, University of Athens, GR-15771 Athens, Greece}
\newcommand{\NTUAthens}{Department of Physics, National Tech. University of Athens, GR-15780 Athens, Greece}
\newcommand{\Benedictine}{Physics Department, Benedictine University, Lisle, Illinois 60532, USA}
\newcommand{\BNL}{Brookhaven National Laboratory, Upton, New York 11973, USA}
\newcommand{\CdF}{APC -- Universit\'{e} Paris 7 Denis Diderot, 10, rue Alice Domon et L\'{e}onie Duquet, F-75205 Paris Cedex 13, France}
\newcommand{\Cleveland}{Cleveland Clinic, Cleveland, Ohio 44195, USA}
\newcommand{\Delhi}{Department of Physics and Astrophysics, University of Delhi, Delhi 110007, India}
\newcommand{\GEHealth}{GE Healthcare, Florence South Carolina 29501, USA}
\newcommand{\Harvard}{Department of Physics, Harvard University, Cambridge, Massachusetts 02138, USA}
\newcommand{\HolyCross}{Holy Cross College, Notre Dame, Indiana 46556, USA}
\newcommand{\IIT}{Physics Division, Illinois Institute of Technology, Chicago, Illinois 60616, USA}
\newcommand{\Indiana}{Indiana University, Bloomington, Indiana 47405, USA}
\newcommand{\ITEP}{High Energy Experimental Physics Department, ITEP, 117218 Moscow, Russia}
\newcommand{\JMU}{Physics Department, James Madison University, Harrisonburg, Virginia 22807, USA}
\newcommand{\LASL}{Nuclear Nonproliferation Division, Threat Reduction Directorate, Los Alamos National Laboratory, Los Alamos, New Mexico 87545, USA}
\newcommand{\Lebedev}{Nuclear Physics Department, Lebedev Physical Institute, Leninsky Prospect 53, 119991 Moscow, Russia}
\newcommand{\LLL}{Lawrence Livermore National Laboratory, Livermore, California 94550, USA}
\newcommand{\MIT}{Lincoln Laboratory, Massachusetts Institute of Technology, Lexington, Massachusetts 02420, USA}
\newcommand{\Minnesota}{University of Minnesota, Minneapolis, Minnesota 55455, USA}
\newcommand{\Crookston}{Math, Science and Technology Department, University of Minnesota -- Crookston, Crookston, Minnesota 56716, USA}
\newcommand{\Duluth}{Department of Physics, University of Minnesota -- Duluth, Duluth, Minnesota 55812, USA}
\newcommand{\Oxford}{Subdepartment of Particle Physics, University of Oxford, Oxford OX1 3RH, United Kingdom}
\newcommand{\Pittsburgh}{Department of Physics and Astronomy, University of Pittsburgh, Pittsburgh, Pennsylvania 15260, USA}
\newcommand{\IHEP}{Institute for High Energy Physics, Protvino, Moscow Region RU-140284, Russia}
\newcommand{\RoyalH}{Physics Department, Royal Holloway, University of London, Egham, Surrey, TW20 0EX, United Kingdom}
\newcommand{\Carolina}{Department of Physics and Astronomy, University of South Carolina, Columbia, South Carolina 29208, USA}
\newcommand{\SLAC}{Stanford Linear Accelerator Center, Stanford, California 94309, USA}
\newcommand{\Stanford}{Department of Physics, Stanford University, Stanford, California 94305, USA}
\newcommand{\StJohnFisher}{Physics Department, St. John Fisher College, Rochester, New York 14618 USA}
\newcommand{\Sussex}{Department of Physics and Astronomy, University of Sussex, Falmer, Brighton BN1 9QH, United Kingdom}
\newcommand{\TexasAM}{Physics Department, Texas A\&M University, College Station, Texas 77843, USA}
\newcommand{\Texas}{Department of Physics, University of Texas, 1 University Station C1600, Austin, Texas 78712, USA}
\newcommand{\TechX}{Tech-X Corporation, Boulder, Colorado 80303, USA}
\newcommand{\Tufts}{Physics Department, Tufts University, Medford, Massachusetts 02155, USA}
\newcommand{\UNICAMP}{Universidade Estadual de Campinas, IF-UNICAMP, CP 6165, 13083-970, Campinas, SP, Brazil}
\newcommand{\USP}{Instituto de F\'{i}sica, Universidade de S\~{a}o Paulo,  CP 66318, 05315-970, S\~{a}o Paulo, SP, Brazil}
\newcommand{\Warsaw}{Department of Physics, Warsaw University, PL-00-681 Warsaw, Poland}
\newcommand{\Washington}{Physics Department, Western Washington University, Bellingham, Washington 98225, USA}
\newcommand{\WandM}{Department of Physics, College of William \& Mary, Williamsburg, Virginia 23187, USA}
\newcommand{\Wisconsin}{Physics Department, University of Wisconsin, Madison, Wisconsin 53706, USA}
\newcommand{\deceased}{Deceased.}

\affiliation{\ANL}
\affiliation{\Athens}
\affiliation{\Benedictine}
\affiliation{\BNL}
\affiliation{\Caltech}
\affiliation{\Cambridge}
\affiliation{\UNICAMP}
\affiliation{\CdF}
\affiliation{\FNAL}
\affiliation{\Harvard}
\affiliation{\IIT}
\affiliation{\Indiana}
\affiliation{\ITEP}
\affiliation{\JMU}
\affiliation{\Lebedev}
\affiliation{\LLL}
\affiliation{\UCL}
\affiliation{\Minnesota}
\affiliation{\Duluth}
\affiliation{\Oxford}
\affiliation{\Pittsburgh}
\affiliation{\RAL}
\affiliation{\USP}
\affiliation{\Carolina}
\affiliation{\Stanford}
\affiliation{\Sussex}
\affiliation{\TexasAM}
\affiliation{\Texas}
\affiliation{\Tufts}
\affiliation{\Warsaw}
\affiliation{\Washington}
\affiliation{\WandM}

\author{P.~Adamson}
\affiliation{\FNAL}

\author{C.~Andreopoulos}
\affiliation{\RAL}

\author{K.~E.~Arms}
\affiliation{\Minnesota}

\author{R.~Armstrong}
\affiliation{\Indiana}

\author{D.~J.~Auty}
\affiliation{\Sussex}


\author{D.~S.~Ayres}
\affiliation{\ANL}

\author{B.~Baller}
\affiliation{\FNAL}


\author{P.~D.~Barnes~Jr.}
\affiliation{\LLL}

\author{G.~Barr}
\affiliation{\Oxford}

\author{W.~L.~Barrett}
\affiliation{\Washington}


% ...

\collaboration{The MINOS Collaboration}
\noaffiliation

\begin{abstract}
This letter reports new results from the MINOS experiment based on
      a two-year exposure to muon neutrinos from the Fermilab NuMI beam.  Our data are consistent with quantum
      mechanical oscillations of neutrino flavor with mass splitting $|\Delta m^2|=(2.43\pm 0.13)\times10^{-3}$~eV$^2$ (68\% confidence level) and mixing angle $\sin^2(2\theta)>0.90$~(90\% confidence level).  Our data disfavor two alternative explanations for the disappearance of neutrinos in flight, namely neutrino decays into lighter particles and quantum decoherence of neutrinos, at the 3.7 and 5.7 standard deviation levels, respectively.
\end{abstract}
\pacs{14.60.Lm, 14.60.Pq, 29.27.-a, 29.30.-h}
\maketitle

Several experiments 
\cite{ref:pdg} have produced compelling evidence of the disappearance of neutrinos of a given lepton flavor.  In previous publications \cite{ref:minosprl}, the 
MINOS 
experiment has also presented evidence for energy-dependent disappearance of muon neutrinos produced by the 
NuMI 
facility at Fermilab.  Based on the number of events, that result provides evidence of the disappearance of $\nu_\mu$ 
at a level of 5.2~standard deviations.  
Such observations support the description of neutrinos via two independent basis sets, the mass and the flavor eigenstates, with the bases related by the 3$\times$3 PMNS matrix\,\cite{ref:pmns}.  
They imply that at least two of the neutrino eigenstates have non-zero mass.  In this letter we present results obtained from a larger dataset than that used in~\cite{ref:minosprl}.  These results provide a precision measurement of the oscillation parameters and further disfavor two other theoretical interpretations of neutrino flavor disappearance \cite{ref:decay,ref:decohere}.

The MINOS detectors \cite{ref:minos} and the NuMI beam line \cite{ref:numi} are described elsewhere. In brief, NuMI is a conventional two-horn-focused neutrino beam with a 675~m long decay tunnel.  The horn current and position of the hadron production target relative to the horns can be configured to produce different $\nu_\mu$ energy spectra.  MINOS consists of two detectors: a 0.98~kt Near Detector (ND) 1.04~km from the NuMI target; and a 5.4~kt Far Detector (FD) 735\,km from the target.   Both are segmented, magnetized calorimeters that permit particle tracking.   The curvature of muons produced in $\nu_\mu +\mbox{Fe}\rightarrow\mu^-+X$ interactions \footnote{Approximately 5\% of $\nu_\mu$ interactions occur in aluminum and scintillator.} is used for energy determination of muons that exit the detector and to distinguish the $\nu_\mu$ component of the beam from the 6\% intrinsic $\overline{\nu}_\mu$ contamination.  The energy of muons contained in the detector is measured via their range.  Oscillations of $\nu_\mu$ into other neutrino flavors result in an energy-dependent depletion of $\nu_\mu$ interactions in the FD relative to the expectation based upon the ND measurement.  

The present letter describes results from data recorded between May 2005, and July 2007.  Over this period, a total of $3.36\times10^{20}$~protons on target (POT) were accumulated for this analysis.  A $1.27\times10^{20}$~POT subset of this exposure (hereafter referred to as Run~I) forms the data set  from Ref~\cite{ref:minosprl}.  In Run~I and for most of the new running period (Run II), the  beam line was configured to enhance $\nu_\mu$ production with energies 1-5~GeV (the low-energy configuration).  An exposure of $0.15\times10^{20}$~POT was accumulated with the beam line configured to enhance the $\nu_\mu$ energy spectrum at 5-10\,GeV (the high-energy configuration).  The Run II data were collected with a replacement target of identical construction due to failure of the motion system of the first target.  The new target was found to be displaced longitudinally $\sim$1~cm relative to the first target, resulting in a 30 MeV shift in the neutrino spectrum.  This effect is incorporated in the Monte Carlo simulation, and the Run I and Run II data sets are analyzed separately to account for this shift.

%

\begin{figure}
\includegraphics[width=2.9in]{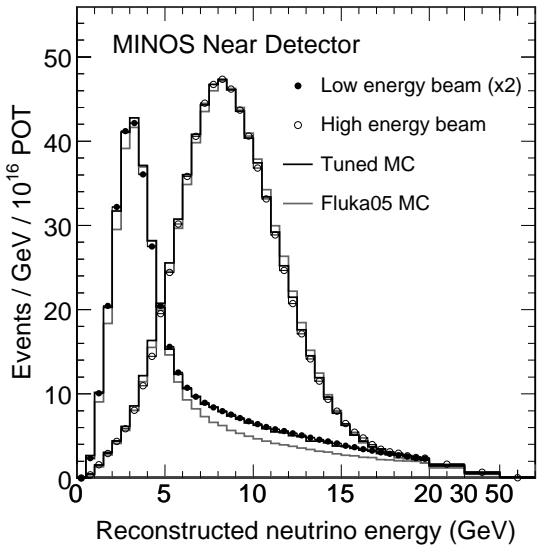}
\caption{\label{fig:tuning} 
Energy spectra in the MINOS ND for two of the nine beam configurations before and after tuning the Monte Carlo simulation to the ND data.  The data combine Run I + Run II.  Both configurations are utilized in the oscillation analysis.}
\end{figure}

%
%
The simulation of neutrino production and detection is accomplished with a model of hadron production in the target using \textsc{Fluka}\,\cite{ref:fluka} and a \textsc{geant3}\,\cite{ref:geant} simulation of the beamline and detector.  \textsc{neugen3.5.5}\,\cite{ref:nugen355}, tuned to data from previous bubble chamber neutrino experiments and experiments with pion beams scattering on iron, is used to model neutrino interactions.  As in our previous analysis, the Monte Carlo (MC) simulation of the neutrino flux was constrained to agree with the neutrino energy spectrum in the ND collected in nine different configurations of the NuMI beam\,\cite{ref:minosprl}, thereby reducing 
uncertainties in the flux prediction at the FD.  Fig.\,\ref{fig:tuning} compares the simulation to the 
ND data acquired in the two configurations used in the oscillation analysis.

Neutrino interactions in the MINOS detectors can either be charged-current, $\nu_\mu+\mbox{Fe}\rightarrow\mu^-+X$, or neutral-current, $\nu_\mu+\mbox{Fe}\rightarrow\nu_\mu +X$.  In this analysis, only the former are used because they identify the interacting neutrino flavor and because the reconstructed energy best measures the full neutrino energy.  
To select charged-current events, we have implemented a new algorithm\,\cite{ref:rustem} based on a multivariate likelihood including four variables that characterize a muon track: the event length; the average pulse height per plane along the track; the transverse energy deposition profile of the track; and the fluctuation of the energy deposited in scintillator strips along the track.
The new selection algorithm, along with a new track-finding algorithm, improve our efficiency to identify and select charged-current interactions in the FD from 75.3\% using the previous selection\,\cite{ref:minosprl} to 81.5\% in the current selection, in the absence of oscillations.  
The new selection reduces the neutral-current contamination in the charged-current sample from 1.8\% in our previous publication to 0.6\% in the present analysis.  The present analysis uses a larger fiducial mass of 4.17 kt in the FD, an increase of 2.9\% over the mass used in \cite{ref:minosprl}.  

The measured energy spectrum at the ND is used to predict the energy spectrum at the FD. 
As in our previous analysis\,\cite{ref:minosprl}, we compute a transfer matrix to correct for $\sim$20\% differences expected in the shape of the energy spectrum in the FD relative to the ND that arise from meson decay kinematics and from beam-line geometry \cite{ref:adam,ref:minosprl}.  We have cross-checked this technique by comparison to other calculations of the FD spectrum\,\cite{ref:minosprl}.

\begin{figure}
\includegraphics[width=3.4in]{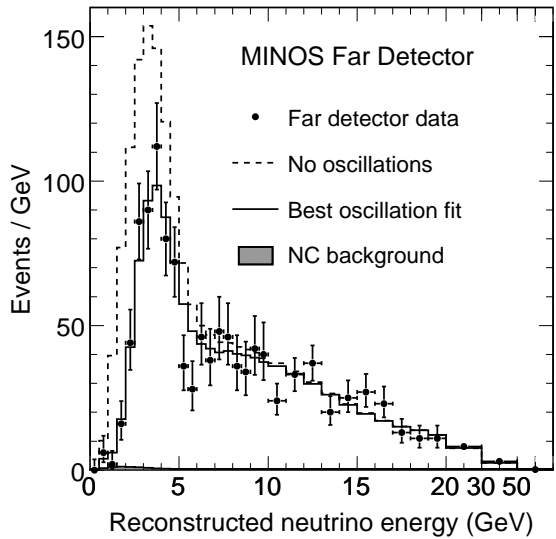}
\caption{\label{fig:spectra} 
Comparison of the FD data (points, with statistical uncertainties) from the low- and high-energy configurations with the predictions for the $\nu_\mu$ energy spectra with and without the effect of oscillations. 
The estimated neutral current (NC) background is indicated.   
}\end{figure}

\begin{figure}
\includegraphics[width=3.2in]{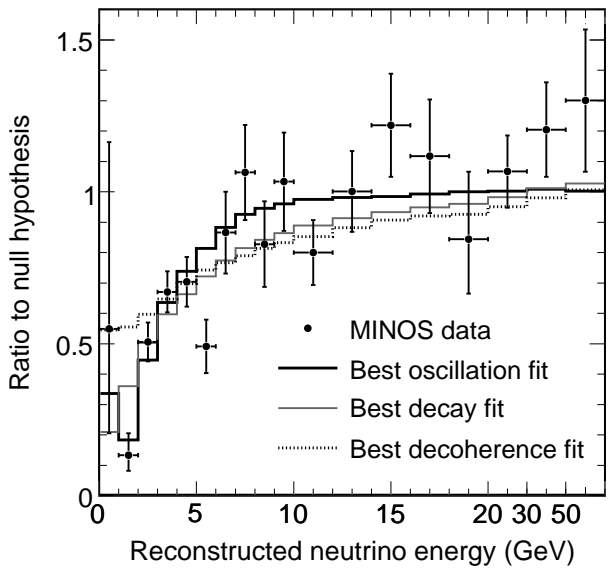}
\caption{\label{fig:ratio} 
Ratio of the FD data and the expected spectrum in the absence of oscillations.  Also shown are the best fit curve to Eq.\,\ref{eq:osc} and the best fit 
 to alternative models of neutrino disappearance\,\cite{ref:decay,ref:decohere}.  For display purposes, the data have been rebinned and the estimated oscillated NC background is subtracted.
}\end{figure}

The FD energy spectra were inspected only after the analysis procedure was finalized and basic data integrity checks were performed.  We observe 848 events in the FD for all energies 0-120\,GeV produced by the NuMI beam, compared to the unoscillated expectation of 1065$\pm$60~(syst.).  In the low energy configuration alone, the number of events observed in the data is 730, to be compared with an expectation of $936\pm53$~(syst.).  The observed energy spectrum of the events from the low- and high-energy datasets is shown along with the predicted spectrum in Fig.\,\ref{fig:spectra} and the ratio of these data to the expected spectrum is shown in Fig.\,\ref{fig:ratio}.
%
%

Under the assumption the observed deficit is due to \numutonutau{}oscillations\,\cite{ref:pdg}, 
a fit is performed to extract the parameters \dmsq{} and \sintwo{}\,\cite{ref:fogli} using the expression
\begin{equation}  P(\nu_{\mu} \rightarrow \nu_{\mu})=1-\rm{sin}^2(2\theta) \rm{sin}^2\left(1.27\Delta {\it m}^2\frac{\it L}{\it E}\right), \label{eq:osc}  \end{equation}
where $L$[km] is the distance from the target, $E$[GeV] is the neutrino energy, and \dmsq{} is measured in \evsq.  The FD data from Run~I, Run~II, and the high-energy run are separately fit to Eq.~\ref{eq:osc}.  
The best-fit parameters minimize the $\chi^2$ expression given in \cite{ref:minosprl}.  
The predicted oscillated spectrum includes the contamination from \nutau{} produced in the oscillation process.


The effects of systematic uncertainties were evaluated by fitting modified MC in place of data.  Table\,\ref{tab:systematics} gives the differences between the fitted values obtained with the modified and unmodified MC.  The largest effects are:  (a) the $\pm10.3\%$ uncertainty in the absolute hadronic energy scale, which is the sum in quadrature of a $\pm5.7$\% error in the calorimeter response to hadrons as derived from test beam measurements \cite{ref:caldet}, a $\pm$2.3\% uncertainty in the energy scale calibration, and a $\pm8.2$\% uncertainty in the simulation of neutrino production of hadrons in iron nuclei; (b) the $\pm$3.3\% relative uncertainty in the hadronic energy scale between the ND and FD; (c) the $\pm$4.0\% uncertainty on the predicted FD event rate which is the sum in quadrature of the uncertainties on the detectors' fiducial mass, event selection efficiency and the POT counting; (d) the $\pm50\%$ uncertainty on the neutral current contamination in the charged current event sample; and (e) the uncertainty on the muon momenta measured via range ($\pm2.0\%$) or curvature ($\pm3.0\%$).  

\begin{table}
\begin{ruledtabular}
\begin{tabular}{lcc}
Uncertainty 							& \dmsq					& \sintwo \\
									&(10$^{-3}$ eV$^2)$         &             \\ \hline
(a)\,Abs hadronic $E$ scale ($\pm$ 10.3\%) 	& 0.052 					& 0.004 \\
(b)\,Rel hadronic $E$ scale ($\pm$  3.3\%)	& 0.027					& 0.006 \\
(c)\,Normalization ($\pm$  4\%)				& 0.081 					& 0.001 \\
(d)\,NC contamination ($\pm$ 50\%)  		& 0.021 					& 0.016 \\
(e)\,$\mu$ momentum (range 2\%, curv 3\%)	& 0.032 					& 0.003 \\
(f)\,$\sigma_\nu(E_\nu<10$~GeV) ($\pm$12\%) & 0.006 					& 0.004 \\
(g)\,Beam flux 							& 0.010 					& 0.000 \\\hline
Total Systematic Uncertainty 				&  0.108  					& 0.018\\
Expected Statistical Uncertainty			& 0.19		                            & 0.09
\end{tabular}
\end{ruledtabular}

\caption{\label{tab:systematics}
Sources of systematic uncertainties in the
measurement of \dmsq{} and \sintwo{}.
The values are the average shifts for varying the parameters in both
directions without imposing the \sintwo$\leq 1$ constraint on the fit. 
Correlations between the systematic 
effects are not taken into account.  The dominant uncertainties are incorporated as nuisance parameters in the fit of our data to Eq.~\ref{eq:osc} so as to reduce their effect on the oscillation parameter measurement (see text). 
}
\end{table}

\begin{figure}[t]
\includegraphics[width=3.1in]{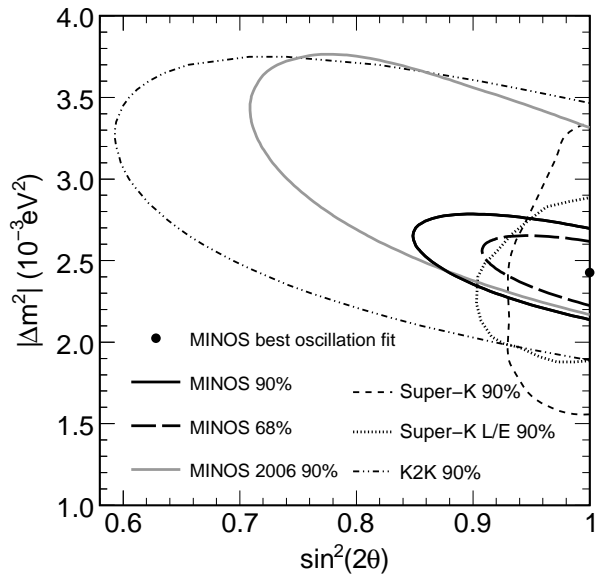}
\caption{\label{fig:contour}  Contours for the oscillation fit to the data in Fig.\,\ref{fig:spectra}, 
including systematic errors.  Also shown are contours from previous experiments\,\cite{ref:osc1,ref:osc5} and our earlier result~\cite{ref:minosprl}.}
\end{figure}

In fitting the data to Eq.\,\ref{eq:osc}, \sintwo{} was constrained to lie in the physical region.  To reduce the effect of the dominant systematic uncertainties in Table\,\ref{tab:systematics} ((a) and (c) for \dmsq{} , and (d) for \sintwo{}) these three systematic uncertainties were included as nuisance parameters in the fit.  The resulting best fit to the neutrino energy spectrum is shown in Fig.~\ref{fig:spectra} and Fig.~\ref{fig:ratio}.  We obtain \dmsq\,=(2.43\,$\pm0.13)\times 10^{-3}$\,\evsq{} and \sintwo\,$>$\,0.95 at 68\% confidence level (C.L.)\,\cite{ref:dof}.   The fit $\chi^2$=90 for 97~degrees of freedom.  The resulting 68\% C.L. ($\Delta$\cs=2.30) and 90\% C.L. ($\Delta$\cs=4.61) intervals for the oscillation parameters \dmsq{} and \sintwo{} are shown in Fig.\,\ref{fig:contour}\,\cite{ref:FC}.   The MC predicts negligible backgrounds of $0.7$ events from cosmic ray muons, and, at the best-fit value for \dmsq{} and \sintwo{}, $2.3$ events from neutrino interactions in the upstream rock, 5.9 neutral current and 1.5 \nutau{} events in the final sample.  If the fit is not constrained to the physical region, \dmsq{}=2.33\,$\times 10^{-3}$\,\evsq{} and \sintwo{}= 1.07, with a 0.6 unit decrease in \cs{}.  Correspondingly, the contours in Fig.~\ref{fig:contour} are smaller than those expected for the present data set.  Our measurement is the most precise determination of the mass splitting \dmsq.

Fig.~\ref{fig:contour} also shows that the previous MINOS result~\cite{ref:minosprl} is in good agreement with the current measurement.  Taken alone, the Run II data yield \dmsq{}=$(2.32^{+0.17}_{-0.16})\times 10^{-3}$\,\evsq{} and \sintwo{}= 1.0, to be compared with (2.57$^{+0.23}_{-0.20})\times10^{-3}$~\evsq{} and \sintwo{}=1.0 from Run~I.  The two results are consistent at 68\% C.L.  
We note that the value of 2.57$\times10^{-3}$~\evsq{} for Run I differs from that quoted in \cite{ref:minosprl}  
because of our improved reconstruction and selection of 
charged-current events and improved MC simulation of neutrino interactions.

We have also fit the FD energy spectra to alternative models that have been proposed to explain the disappearance of neutrinos in flight, namely, the decay of neutrinos to lighter particles (Eq.~13 of \cite{ref:decay}), and the decoherence of the neutrino's quantum-mechanical wave packet (Eq.~5 of \cite{ref:decohere}).  Fig.\,\ref{fig:ratio} shows the ratios of the energy spectra arising from our best fits to these alternative models to the prediction of the FD spectrum in the absence of $\nu_\mu$ disappearance.  The $\chi^2$ for the best fit to the decay model is 104/97~d.o.f., while that for the decoherence model is 123/97~d.o.f.  Given the $\Delta \chi^2=14$ and 33 of these two models relative to the oscillation hypothesis, these models are disfavored with respect to the oscillation hypothesis at the 3.7 and 5.7 standard-deviation levels.

In summary, we have presented updated measurements of neutrino oscillation parameters from the MINOS experiment.  Based upon an exposure of $3.36\times10^{20}$~POT from the NuMI beam, we obtain $|\Delta m^2|=(2.43\pm 0.13)\times10^{-3}$~eV$^2$ (68\% C.L.) and mixing angle $\sin^2(2\theta)>0.90$~(90\% C.L.).  As the dataset presented here includes the subset analyzed in \cite{ref:minosprl}, these results supersede our previous publication.  Our data disfavor two alternative explanations for disappearance of neutrinos in flight, namely neutrino decays~\cite{ref:decay} into lighter particles and quantum decoherence of neutrinos~\cite{ref:decohere} at the 3.7 and 5.7 standard-deviation level, respectively.  

This work was supported by the US DOE; the UK STFC; the US NSF; the State and University of Minnesota; the University of Athens, Greece; and Brazil's FAPESP and CNPq.  We are grateful to the Minnesota Department of Natural Resources,
the crew of the Soudan Underground Laboratory, and the staff of Fermilab for their contribution to this effort.


\begin{thebibliography}{99}
\bibitem{ref:pdg} W.-M. Yao {\it et al.}, J. Phys. G 33, 1 (2006).
\bibitem{ref:minosprl} D.G. Michael {\it et al.}, Phys. Rev. Lett. 97, 191801 (2006);
P. Adamson et al, Phys. Rev. D77, 072002 (2008).
\bibitem{ref:pmns}B. Pontecorvo, JETP 34, 172 (1958).  
Z. Maki, M. Nakagawa, and S. Sakata, Prog. Theor. Phys. 28, 870 (1962).
\bibitem{ref:decay} V. Barger {\it et al.},  Phys. Rev. Lett. 82 2640 (1999).                        
\bibitem{ref:decohere} G. L. Fogli {\it et al.},  Phys. Rev. D67 093006 (2003).                             
\bibitem{ref:minos} D.G. Michael {\it et al.}, submitted to Nucl. Instr. and Meth.
\bibitem{ref:numi} S. Kopp, Proc. 2005 IEEE Part. Accel. Conf., May 2005, Fermilab-Conf-05-093-AD and {\tt arXiv:physics/0508001}. 
\bibitem{ref:fluka}A. Fasso {\it et al.}, CERN-2005-10, INFN/TC\_05/11, SLAC-R-773 (2005).
\bibitem{ref:geant}R. Brun et al., CERN Program Library W5013 (1984).
\bibitem{ref:nugen355} H. Gallagher, Nucl. Phys. B (Proc. Suppl.) 112, 188 (2002); update at {\tt arXiv:0806.2119} (2008).
\bibitem{ref:rustem} R. Ospanov, PhD Dissertation, UT-Austin, 2008
\bibitem{ref:adam}M. Szleper and A. Para, hep-ex/0110001.
\bibitem{ref:fogli}{The experiment measures an unresolved 
mixture of \dmsqone{} and \dmsqtwo{}, which we refer to as \dmsq{} for brevity.  The parameter
\sintwo{} is likewise an admixture.
For further discussion see G. L. Fogli {\it et al.}, Prog. Part. Nucl. Phys. 57, 742 (2006).}
\bibitem{ref:caldet}P. Adamson {\it et al.}, Nucl. Inst. \& Meth. A556, 119 (2006).
\bibitem{ref:dof}{Although the contours in Fig.\,\ref{fig:contour} 
are calculated with two degrees of freedom (d.o.f.), 
the parameter errors are calculated
with only one d.o.f.\,as in \cite{ref:pdg} using $\Delta\chi^2=1$ and 2.71, respectively.}
\bibitem{ref:FC} {The effect of the constraint to the physical region was investigated using the unified
approach of G.J. Feldman and R.D. Cousins, Phys. Rev. D57, 3873, (1998), which gave slightly smaller confidence intervals}.
\bibitem{ref:osc1}Y. Ashie {\it et al.}, Phys. Rev. Lett. 93, 101801 (2004); Phys. Rev. D71, 112005 (2005).
\bibitem{ref:osc5}M.H. Ahn {\it et al.}, Phys Rev D 74 072003 (2006).

\end{thebibliography}
\end{document}